\documentstyle[12pt,boxedminipage,epsfig,wrapfig,floatflt,shadow]{article}

\begin{document}

\title{Categorization of Quantum Mechanics Problems by Professors and Students}

\author{Shih-Yin Lin and Chandralekha Singh,\\
  Department of Physics and Astronomy,\\
 University of Pittsburgh, Pittsburgh, PA, 15260}
\date{ }

\maketitle

\begin{abstract}
We discuss the categorization of 20 quantum mechanics problems by physics professors and undergraduate students from two honors-level
quantum mechanics courses. Professors and students were asked to categorize the problems based upon similarity of
solution. We also had individual discussions with professors who categorized the problems.
Faculty members' categorizations were overall rated higher than those of students by three faculty members who evaluated
all of the categorizations. The categories created by faculty members were more diverse compared to the
categories they created for a set of introductory mechanics problems. Some faculty members noted that the categorization
of introductory physics problems often involves identifying fundamental principles relevant for the problem, whereas in upper-level
undergraduate quantum mechanics problems, it mainly involves
identifying concepts and procedures required to solve the problem. Moreover, physics faculty members who evaluated
others' categorizations expressed that the task was very challenging and they sometimes found another person's categorization to be better than their own.
They also rated some concrete categories such as ``hydrogen atom" or ``simple harmonic oscillator"
higher than other concrete categories such as ``infinite square well" or ``free particle".
\end{abstract}

\section{Introduction}
\vspace*{-.015in}

A crucial difference between the problem solving strategies used by experts in physics
and beginning students lies in the interplay
between how their knowledge is organized and how it is retrieved to solve problems~\cite{chi3,hardiman,reif1,physed,reif2,chandra}.
Categorizing or grouping together problems based upon similarity
of solution can give a glimpse of the ``pattern" an individual sees in a problem while contemplating how to solve it~\cite{chi3}.
In a classic study by Chi et al.~\cite{chi3}, a categorization task was used to assess introductory physics students' level of expertise in physics.
In Chi's study~\cite{chi3}, eight introductory physics students were asked to group together introductory mechanics
problems into categories based upon similarity of solution.
They found that, unlike experts (physics graduate students in their study) who categorized them 
based on the physical principles required to solve them, introductory students 
categorized problems involving inclined planes in one category and pulleys in a separate category~\cite{chi3}.
Previously, we conducted a categorization study in which 7 professors, 21 physics graduate students and more than a hundred introductory physics students 
in a classroom environment were asked to group together introductory physics problems based upon similarity of solution~\cite{chandra}.
We found that the professors significantly outperformed both the graduate students and introductory physics students in grouping together
problems based upon the physics principles involved rather than basing the grouping of the problems on the surface features of the problems 
and they 
created very similar categories~\cite{chandra}. The graduate students performed better than the introductory physics students in the categorization
task. However, there is a large overlap in the performance of graduate students and introductory
students in the calculus-based courses on the categorization of introductory physics problems into groups based upon the fundamental principles
of physics required to solve the problems~\cite{chandra}.

While learning introductory physics is challenging, learning quantum mechanics is perhaps even more so~\cite{phystoday,jolly,my1,katja,theme,quantum2,my2,my3,my4,my5,my7,my9,quantum}.
Unlike classical mechanics, we do not have direct experience
with the microscopic quantum world. Also, quantum mechanics has an abstract theoretical framework in which the most fundamental
equation, the Time-Dependent Schroedinger Equation (TDSE), describes the time evolution of the wave function or the state of a quantum system
according to the Hamiltonian of the system.
This wave function is in general complex and does not directly represent a physical entity. However, the wave function at a given time can be
used to calculate the probability of measuring a particular value for a given physical observable associated with the system. For example, the absolute square of the
wave function in position-space gives the probability density.
Since the TDSE does not describe the evolution or motion of a physical entity, unlike Newton's second law, the modeling of the microscopic world in
quantum mechanics is generally more abstract than the modeling of the macroscopic world in classical mechanics.

The conceptual framework of quantum mechanics is often counter-intuitive to our everyday experiences.
According to quantum theory, the position, momentum, energy and other observables for a quantum mechanical entity
are in general not well-defined. We can
only predict the probability of measuring different values based upon the wave function when a measurement is performed.
This probabilistic interpretation of quantum mechanics, which even Einstein found disconcerting, is challenging for students.
Moreover, according to the Copenhagen interpretation of quantum mechanics, which is widely taught to students, measurement of a physical
observable ``collapses" the wave function into an eigenstate
of the operator corresponding to the observable
measured. Thus, the usual time evolution of the system according to the TDSE is treated differently from measurement processes.
Students often have difficulty with this notion of an instantaneous change or ``collapse" of the wave function during the measurement [13].
The proper way to interpret quantum mechanics is still the subject of debate, making the subject even more challenging for physics instructors.

Here, we discuss a study in which 22 physics juniors and seniors 
in two undergraduate quantum mechanics courses and six physics faculty members (professors)
were asked to categorize 20 quantum mechanics problems based upon similarity of solution. We also interviewed some faculty members concerning 
issues related to categorization of quantum mechanics problems. 
All but one faculty member had taught an upper-level undergraduate or graduate level quantum mechanics course. 
The faculty member who had not taught quantum mechanics regularly teaches other physics graduate ``core" courses
including electricity and magnetism and statistical mechanics. All undergraduate students in the upper-level quantum mechanics
classes (12 and 10 students in the two classes 
who were present on the day the categorization task was given as a quiz) participated. The students were given 35-40 minutes to perform
the categorization. The faculty members performed the categorization at a time convenient to them. Except for the faculty member who
had not taught quantum mechanics and took longer to categorize the problems, other professors noted that it took them
less than 30 minutes to perform the categorizations.

The 20 problems to be categorized (given in the Appendix)
were adapted from the problems found among the end of the chapter exercises in commonly used upper-level
undergraduate quantum mechanics textbooks. All those who performed the categorization were provided
the instructions given at the beginning of the Appendix. The sheet
on which individuals were asked to perform the categorization of problems had three columns. 
In the first column, they were asked to place their
own {}``category name\char`\"{} for each category (in other words, they had to come up with their own category names), in the second
column, they had to place a description of the category that explains
why those problems can be grouped together; in the third column,
they had to list the problem numbers for the problems that should be placed in that category.
We note that for solving a problem, more than one approach
may be useful. The instruction for the categorization explicitly noted that a problem could be placed in more than one category. 

The goal was to investigate differences in categorization
by faculty members and students and whether there are major differences in the ways in which individuals in each group categorize quantum mechanics 
problems. This study was partly inspired by the fact that a physics faculty member 
who was teaching advanced 
undergraduate quantum mechanics in a previous semester had given a take-home exam in which one problem asked students to find the wave function of a 
free particle after a time $t$ given the initial wave function (which was a Gaussian). Two students approached the faculty member complaining that this 
material was not covered in the class. The faculty member pointed out to them that he had discussed in the class how to find the wave function 
after a time $t$ given an initial wave function in the context of a problem involving an infinite square well. But the students insisted that, while the 
time-development of the wave function may have been discussed in the context of an infinite square well, it was not discussed in the context of a free particle. It 
appears that the two students did not categorize the time-development issues for the infinite square well and the free particle in the same category. 
They did not realize that a solution procedure very similar to what they had learned in the context of the time-development of the wave function for an 
infinite square well should be applicable to the free particle case except they must use the energy eigenstates and eigenvalues corresponding to the 
free particle and replace the discrete sum over energy levels for an infinite square well by an integral since the energy levels for a free particle 
are continuous. This difficulty in discerning that the same concepts and procedures should be applicable in both contexts is similar to the
difficulty introductory students have in discerning that the same principle is applicable in two problems that have different contexts.

\vspace*{-.231in}
\section{Scoring of Categorization}
\vspace*{-.11in}

We note that each individual who categorized the problems had to come up with his/her own category names
and justify why each problem should be placed in a particular category.
The 20 questions for categorization were such that the ``context" in four of them was the hydrogen atom, the harmonic oscillator, the infinite square well,
and the free particle (see the Appendix). Three of the problems were related to the spin angular momentum and one was about the Dirac delta function. 
Within these different contexts, there were questions about the time evolution of wave function, time dependence of expectation value, measurement of 
physical observables, expectation value including uncertainty, commutation relation between different components of the spin angular momentum etc. 
As noted earlier, we wanted to investigate if the questions were grouped together based upon the physics concepts and procedures required
for solving them or the ``surface features" of the problems such as the contexts used.
For problems related to the time-dependence of wave function or time-dependence of expectation value, we wanted to investigate if the faculty members
and students categorized problems involving the stationary states differently from those involving the non-stationary states.

We find that the categorizations of a problem performed by the students were diverse and they seldom placed a problem in more than
one category although they were explicitly told they could do so if they wish. Moreover, the faculty members often used a diverse set of categories
unlike the highly uniform categorization by faculty members for introductory physics problems~\cite{chandra}.

To analyze the quality of categories created by the professors and the students quantitatively, we placed each category created by each individual into a matrix which consisted of 
problem numbers along the columns and categories along the rows. A ``1" was assigned if the problem appeared in the given 
category and a ``0" was assigned if the opposite was true. Categories that were very similar were combined, e.g., ``time-dependence of wave function", 
``time-development of wave function" or ``dynamics of wave function" were combined into a single category. 
In order to score the categorizations by students and faculty members, three faculty members (a subset of those who had categorized the problems 
themselves) were recruited. They were given the categorizations by students and faculty in the matrix form we had created 
(without identifiers and with the categorizations by the faculty and students jumbled up). For example, all the different categories created by different 
individuals for problem (1) were placed one after the other to aid faculty members who were scoring the categorizations. 

For each question, the three faculty members doing the scoring were advised to read the question, think about how they would categorize it and then evaluate 
and score everybody's categorization.
They were asked to evaluate whether 
each of the categories created by an individual should be considered {}``good\char`\"{} (assigned a score of 2), {}``moderate\char`\"{} (assigned
a score of 1), or {}``poor\char`\"{} (assigned a score of zero). We note that if all three faculty members scored a particular problem for an 
individual as ``good", the score of that individual on that problem will be 6 (maximum possible). If one faculty member 
scored it as ``good" but the other two scored it as ``medium", the score of that individual on that problem will be 4. 

\vspace*{-.16in}
\section{Results}
\vspace*{-.06in}

Each of the 22 students and 6 faculty members categorized 20 problems. 
Faculty members often placed a problem in more than one category. 
As noted earlier, some of the categories created for a problem by more than one individual were the same or similar.
Several categories that were similar were combined into a single category.
All the three faculty members
noted that evaluating and scoring other people's categorization was a very challenging task and required intense focus. One faculty member noted
that it took him several hours to complete the scoring. Moreover, two of the faculty members who evaluated everybody's categorization noted 
that they would prefer not to use the terms ``good" or ``poor" for judging the categories although some categories were better than others.
The faculty members who scored others' categorizations also noted that 
sometimes they liked the categorizations of a problem by others much more than their own.
Interestingly, in our earlier studies with introductory physics categorization, we had asked three faculty members to evaluate the
categorizations of a subset of randomly selected individuals (in that case we did not ask them to score all categorizations because the introductory
physics classes had several hundred students)~\cite{chandra}. In scoring introductory physics categorizations, faculty members
were not hesitant in
calling the categories good/poor and they did not say that the task was challenging~\cite{chandra}. They also never said that they preferred other's categorizations
of a problem more than their own perhaps because there was a great conformity in faculty categorizations (which 
were based upon physics principles such as 
the conservation of mechanical energy, conservation of momentum, conservation of angular momentum, Newton's second law etc.)~\cite{chandra}.

Table 1 shows examples of category names for each question divided into three groups with a score of ``5 or 6", ``3 or 4" or ``less than 3". 
With each category name, many faculty members and students provided an explanation justifying why certain problems should be placed in that
category. 
Inspection of Table 1 shows that the categories that obtained a total score of less than 3 (out of 6) included both concrete and abstract
categories. For example, ``change in basis" for problem (1), ``commutation relation" for problem (12), ``matrix element" for problem (17),
``rotation group" for problem (19) etc. are abstract categories that received a score of less than 3. On the other hand, ``infinite square
well" for problems (4), (12), (16) and (18) and ``free particle" for problems (3), (7) and (10) etc. are examples of  concrete categories 
that received a score of less than 3. 

Figure 1 shows a histogram of the percentage of people (students or faculty) vs. percentage of problems with a score of $50\%$ of better 
(at least 3 out of 6) and Figure 2 shows a histogram of the percentage of people vs.
average score on the categorization task out of a maximum of 6 (averaged over all problems). 
We note that what one faculty member scored as ``good" was often scored as ``medium" by another.
While three of the six faculty members who categorized the problems were recruited to score all of the categories
by all faculty members and students, the average score of the three faculty members who scored all problems was lower than those of the other three faculty
members who did not score the categories. Also, faculty members who scored the categorizations explicitly noted that they sometimes preferred
others categorizations more than their own. Thus, we do not believe that the faculty members who scored everybody's categorizations
were partial to their own categories.
It is interesting to note that the faculty member who had never taught quantum mechanics (but had taught statistical mechanics and electricity
and magnetism at the graduate level) performed slightly better on average (though not statistically significant) than the faculty members
who scored the categorizations. In fact, the faculty member who had never taught quantum mechanics but performed the categorization
commented that he would like to teach quantum mechanics
but was not assigned that course despite asking for it. He added that the main reason was that
many other faculty members wanted to teach quantum mechanics but they did not want to teach the other graduate level courses that he was assigned.

Figure 1 shows that the categorizations by faculty members 
were rated higher overall than those by students, despite the diversity in faculty responses. 
We find that the faculty members were more likely to categorize the problems based upon the procedures and concepts required to solve the problems
rather than the contexts involved.  
But Figure 2 shows that none of the faculty members had an average score of 5-6 on the categorization task implying none of the 
faculty member placed all 20 problems in categories that were considered uniformly excellent (although their categories were on average better than
those of the students). 
Faculty members sometimes categorized problems based upon the contexts used, e.g., hydrogen atom, simple harmonic oscillator, angular momentum etc.
However, most of the time when they did such categorizations, they also categorized the same problems in other categories which 
were based upon the procedures for solving the problems. 
They were also more likely than students to make
use of the nuances in the questions to group problems, e.g., whether the system was in a stationary state in
order to categorize problems involving the time-dependence of wave function or the time-dependence of expectation value.

The same category was sometimes assigned different scores for different questions depending upon whether the faculty members who scored them felt
they were appropriate categories for those questions.
For example, for question (15), the category ``stationary state" obtained a score of at least 3 because the faculty members felt that it was relevant for
determining the expectation value of momentum and for explaining whether it should depend on time. On the other hand, for question (20), 
``stationary state" category obtained a score less than 3 (see the Appendix) because it was not considered relevant for finding the possible
values of energy after the measurement of the distance of the electron from the nucleus.

Faculty members who scored the categories were careful to distinguish between the categories ``uncertainty" and ``uncertainty principle" (or uncertainty
relation). For example, in questions (10) and (17), ``expectation value and uncertainty" category obtained an average score of 5 or 6 whereas
``uncertainty principle" or ``uncertainty relation" obtained a score of less than 3.
Individual discussions with the faculty who scored the categorization suggests that 
they saw a clear distinction between these categories. In particular, they asserted that calculating the standard deviation $\sigma_x$ 
was about calculating the uncertainty in position but 
it was not about ``uncertainty principle" or ``uncertainty relation". The question did not ask whether
the product of the uncertainties in position and momentum is greater than or equal to $\hbar/2$.

The overall scores (by the three faculty members who evaluated all of the categorizations) on concrete or context-based categories such as
``hydrogen atom" or ``harmonic oscillator" were higher than other concrete categories such as ``infinite square well" or ``free particle", (where
four questions out of 20 given in the categorization task belong to each of these four systems as noted earlier). 
Discussions with individual faculty suggest that
they have a notion of a canonical quantum system that they use for thinking about concepts and to help clarify ideas about quantum mechanics.
``Hydrogen atom" and ``harmonic oscillator" fit their notion of canonical quantum systems. One faculty member explicitly noted that the hydrogen
atom and harmonic oscillator are quintessential in quantum mechanics. He added that the hydrogen atom embodies many essential features of other complex
quantum systems but is exactly soluble and widely applicable. Similarly, the harmonic oscillator is used as a model to understand diverse quantum systems
such as molecular excitations and quantum optics.
Such explanations about why the average score for ``hydrogen atom" as a category was at least $50\%$ (3 out of 6 including the scores
of all the three faculty members who evaluated the responses) for three of 
the four questions that related to the hydrogen atom but was not $50\%$ for any of the four questions related to the infinite square well
shed some light on why the faculty do not view all ``concrete" categories on the same footing. 

As noted earlier, most of the time when faculty members placed problems in a category involving context,
such as ``hydrogen atom" or ``simple harmonic oscillator",
they also placed the same problem in another category based on the procedure involved in solving the problem.
But sometimes they placed some of the problems only in concrete categories. 
For example, one faculty member grouped some problems about hydrogen atom in ``hydrogen atom" category or in a category
based upon the procedure for solving the problems, e.g., ``measurement" or ``time evolution
of wave function" or in both these types of categories. During individual discussions, these faculty members were asked why their choices 
were more context-based in some of their groupings 
and more focused on the procedures and concepts to solve the problems for creating other categories. 
In response, some faculty members reasoned that they were perhaps using lenses with different ``zoom factors" for categorizing different problems. 
They noted that the categorization task was challenging and they sometimes zoomed in and out while categorizing different problems focusing
on the contexts or the procedures for solving them.
Faculty members who scored the categorizations also noted that while scoring others' categorizations they realized that there were many different ways to 
categorize the problems and sometimes others' categorizations were better than their own.

The faculty members were reminded during the individual discussions that while categorizing introductory physics problems, faculty always scored
``inclined plane category", ``cliff category" or ``spring category" as poor categories explaining that they were based on the ``surface
features" of the problems rather than the ``deep" features (fundamental principles of physics required to solve them).
They were asked to comment on whether making categories such as ``angular momentum" or ``hydrogen atom" was also based on the ``surface features" 
of the problems rather than the procedures relevant for solving the problems. In response to such questions, faculty members often noted that while 
these categories were less directly related to the procedure for solving the problems, they were hesitant to call them ``poor" categories. 
They noted, e.g., that the knowledge about the hydrogen atom is relevant for solving the problems involving hydrogen atom even though that knowledge 
alone may not be the 
central component of how to set up the solution of the problem. For example, questions (11) and (14) in the Appendix are about the hydrogen atom 
in a linear superposition of stationary states. In question (11), the knowledge that the expectation value of an operator corresponding to a 
physical observable which does not commute with the Hamiltonian depends on time in a non-stationary state is relevant to solve the problem. 
Similarly, in question (14), knowledge about the time-dependence of wave function in a non-stationary state is relevant for solving the problem.
Simply categorizing these problems in the ``hydrogen atom category" does not indicate whether the individual knows the procedure for solving the 
problem. While the faculty members agreed that some of these concrete categories may not be the best way to categorize the problems,
they sometimes scored some of these context-based categories (even if they did not give an indication of the procedures for
solving the problems) as ``1" instead of ``0" (but rarely gave it a score of ``2"). As shown in Table 1, ``angular momentum" for questions (1) and 
(2), harmonic oscillator for questions (5) and (6), hydrogen atom for questions (8), (14) and (20) are examples of such context-based categories
that were judged favorably.

Individual discussions with faculty members suggest that some felt that the structure of knowledge in quantum mechanics 
is more complex than that in introductory physics.
Moreover, the complexity of knowledge structure in quantum mechanics is due to both the requisite conceptual and mathematical knowledge. 
This complexity may make it difficult for everybody 
to focus on the same aspects of solution when asked to categorize
(although there are often underlying relations in faculty categorizations). 
One possible implication is that the way concepts are emphasized in a quantum mechanics course
may differ based upon the ``patterns" that appear to be most central to the faculty member teaching the course.
For example, one faculty member may emphasize the conceptual aspects while the other may emphasize the mathematical aspects.

During individual discussions, faculty members were asked if they were surprised that the categories in which a problem was placed
by different faculty members were not always similar and some faculty came up with categories that were more abstract than others.
They were also asked to comment on the fact that the faculty members who scored the categorizations gave low scores not only to the concrete categories
but also to some abstract categories. For example, as noted earlier, ``matrix element" for question (17) and ``rotation group" 
for question (19) received a cumulative score of less than 3. 
In response to these questions, faculty members asserted that they were not very surprised about these
because they felt that how one teaches quantum mechanics and 
how abstractly or concretely one presents the material depends strongly on the instructor.
During discussions, several faculty members pointed out that if one takes a look at the quantum mechanics textbooks, he/she will realize that 
the textbooks are laid out very differently and emphasize different things. Some faculty members mentioned that some undergraduate textbooks do not emphasize the postulates of quantum mechanics. Also, the postulates in different textbooks are not identical (e.g., only some of the textbooks list
the Time-dependent Schroedinger equation as a postulate). Some textbooks
are hesitant to mention the ``collapse" of the wave function during measurement while others discuss these issues in detail. They also mentioned that
some textbooks start with the infinite dimensional vector space while others start with the quantum mechanics of a spin-half particle.
The proponents of the spin-half first believe that it provides a simple two dimensional vector space to teach the foundations
of quantum mechanics whereas those
who discuss, e.g., the infinite square well, first believe that spin is too abstract and continuity with the topics covered in
the earlier courses is important.
The extent to which symmetry ideas are emphasized and the conservation laws derived from them also varies in the undergraduate textbooks.
Discussions suggest that most faculty members believed that if there is no agreement on the basic issues about teaching undergraduate quantum mechanics,
the differences in how the faculty members categorize problems, teach their courses and what they emphasize is perhaps expected.

Another common theme that emerged is that categorization
of introductory physics problems involves identifying fundamental principles relevant for the problems, whereas in the upper-level
undergraduate quantum mechanics problems, it mainly involves
identifying concepts and procedures, because problem solving in such a course is tied to conceptual and procedural knowledge.
Some faculty members asserted that the fundamental principles of physics such a conservation of energy and conservation
of momentum are important even for understanding quantum processes. However, the application of fundamental principles to
quantum processes is not typically the focus of an upper-level undergraduate course. For example, one faculty member noted that for understanding
the properties of a solid using neutron scattering, one will have to carefully account for the conservation of energy and momentum but
questions involving these topics are typically not common in an undergraduate quantum mechanics course. He added that if such questions
were given in the categorization task, there may be more uniformity in the faculty responses. 

\vspace*{-.24in}
\section{Summary}
\vspace*{-.13in}

The categorization of problems by students in a quantum mechanics course can be a useful tool for understanding the patterns students
see in a problem when contemplating how to solve it. 
Even in the context of quantum mechanics problems, professors overall scored higher than students in grouping together problems
based on similarity of solutions. 

However, unlike the categorization of introductory physics problems, in which
professors' categorizations are generally uniform, their categorizations were more varied in the context of quantum mechanics.
The diversity of categories created for quantum mechanics may partly be due to the fact that the solution to a typical quantum mechanics
problem in an upper-level quantum mechanics course typically requires the knowledge of requisite concepts and procedures. 
On the other hand, categorization in introductory physics is typically based on the fundamental principles of physics.
Faculty members noted that the fundamental principles, e.g., conservation laws, are also important in understanding quantum 
processes but they are not the focus of an upper-level undergraduate quantum mechanics course.
Some faculty members created more abstract categories than others. It will be useful to investigate how different is the
teaching emphasis of faculty members in a quantum mechanics course depending upon the types of categories they created.

\section{Acknowledgments}
We thank all faculty and students who participated in this study for their help.
This work is supported by the National Science Foundation (PHY-0653129 and 055434).

\pagebreak

\pagebreak

\begin{center}
{\bf Appendix: Categorization Questions} 
\end{center}
$\bullet$ {Your task is to group the 20 problems below into various groups based upon similarity of solution on the sheet of paper provided. 
You can create as many categories as you wish. The grouping of problems should NOT be in terms of
``easy problems", ``medium difficulty problems" and ``difficult problems" but rather it should be based upon the features
and characteristics of the problems that make them similar.  A problem can be placed in more than one group created by you.
Please provide a brief explanation for why you placed a set of questions in a particular group. You need NOT solve any problems.}\\

\noindent
{The first TWO questions refer to the following system:
An electron is in an external magnetic field B which is pointing
in the z direction. The Hamiltonian for the electron spin is given by $\hat H=-\gamma B \hat S_z$ where $\gamma$ is the
gyromagnetic ratio and $\hat S_z$ is the z component of the spin angular momentum operator.\\
}

\begin{enumerate}
\item
If the electron is initially in an eigenstate of $\hat S_x$,
does the expectation value of $\hat S_x$ depend on time?  Justify your answer.

\item
If the electron is initially in an eigenstate of $\hat S_z$, does the expectation value of $\hat S_x$ depend on time?
Justify your answer.

\item
A free particle has the initial wave function
$\Psi(x, t = 0) = Ae^{-ax^2}e^{i k_0 x}$
where $A$, $a$, and $k_0$ are constants ($a$ and $k_0$ are real and positive).
Find $|\Psi(x,t)|^2$. 

\item
A particle in an infinite square well ($0\le x \le a$) has the initial wave function $\psi(x,0)=A x (a-x)$.
Find the uncertainty in position and momentum. 

\item
In the ground state of the harmonic oscillator, what are the expectation values of position, momentum and energy?
Do these expectation values depend on time?

\item
A particle is in the first excited state of a harmonic oscillator potential. Without any calculations, explain
what the expectation value of momentum is and whether it should depend on time.

\item
A free particle has the initial wave function
$\Psi(x, t = 0) = A e^{ik_0 x}$
where $A$, and $k_0$ are constants ($k_0$ is real and positive).
Find $|\Psi(x,t)|^2$. 

\item
An electron is in the ground state of a hydrogen atom.
Find the uncertainty in the energy and the $z$ component of angular momentum.

\item
Make a qualitative sketch of a Dirac delta function $\delta (x)$. Then, make a qualitative sketch of the absolute
value of the Fourier transform
of $\delta (x)$. Label the axes appropriately for each plot.

\item
A free particle has the initial wave function
$\Psi(x, t = 0) = Ae^{-ax^2}e^{i k_0 x}$
where $A$, $a$, and $k_0$ are constants ($a$ and $k_0$ are real and positive).
Find $\langle x\rangle$, $\langle p\rangle$, $\langle x^2\rangle$, $\langle p^2\rangle$, 
$\sigma_x=\sqrt{\langle x^2\rangle- \langle x\rangle^2}$, $\sigma_p=\sqrt{\langle p^2\rangle-\langle p\rangle^2}$.

\item
An electron in a hydrogen atom is in a linear superposition of the first and third excited states. Does the expectation
value of its kinetic energy depend on time?

\item
Suppose that the measurement of the position of a particle in an infinite square well ($0\le x \le a$) yields the value $x=a/2$ at the center of the well.
Show that if energy is measured immediately after the position measurement, it is equally probable to find the particle in any odd-energy
stationary state.

\item
An electron is in a linear combination of the ground and fourth excited states in a harmonic oscillator potential.
A measurement of energy is performed and then followed by a measurement of position. What can you say about the possible
results for the energy and position measurements?

\item 
An electron in a hydrogen atom is in a linear superposition of the first and third excited states.
Find the wave function after time $t$.

\item
A particle is in the third excited state of a harmonic oscillator potential. Without any calculations, explain what the 
expectation value of momentum is and whether it should depend on time.

\item
A particle in an infinite square well ($0\le x \le a$) has the initial wave function $\psi(x,0)=A x (a-x)$.
Without normalizing the wave function, find $\psi(x,t)$.

\item
A free particle has the initial wave function
$\Psi(x, t = 0) = A e^{ik_0 x}$
where $A$, and $k_0$ are constants ($k_0$ is real and positive).
Find $\langle x\rangle$, $\langle p\rangle$, $\langle x^2\rangle$, $\langle p^2\rangle$, 
$\sigma_x=\sqrt{\langle x^2\rangle- \langle x\rangle^2}$, $\sigma_p=\sqrt{\langle p^2\rangle-\langle p\rangle^2}$.

\item
A particle is initially in a linear combination of the ground state and the first excited state of an infinite square well.
Without any calculations, explain whether the expectation value of position should depend on time.

\item
What is the commutation relation [$\hat S_x,\hat S_y$]?

\item
A hydrogen atom is in the first excited state. You measure the distance of the electron from the nucleus first and then
measure energy. Describe the possible values of energy you may measure.
\end{enumerate}

\begin{table}[h]
\centering
\begin{tabular}[t]{|c|c|c|c|}
\hline
Q$\#$& Scores of 5 or 6& Scores of 3 or 4& Scores less than 3 \\[0.5 ex]
\hline \hline
1/2&time dependence of EV/ &eigenvalue$\&$function/angular&Stern-Gerlach/change in basis/  \\[0.5 ex]
&stationary state&momentum/Larmor precession&charged particle in mag. field  \\[0.5 ex]
\hline
3&time evolution of wavefunction&time dependency, evolution&superposition/free particle\\[0.5 ex]
\hline
4& EV/&measurement, observables&infinite square well/\\[0.5 ex]
& EV and uncertainty&and uncertainty relations&$\Psi(x,t)$ manipulations\\[0.5 ex]
\hline
5& EV/eigenstate/&simple harmonic oscillator/&math/\\[0.5 ex]
&time dependence of EV&operator properties/$\Psi(x,t)$&little concept\\[0.5 ex]
\hline
6&time dependence of EV/&simple harmonic oscillator/&matrix element/$\Psi(x,t)$/\\[0.5 ex]
& symmetry argument&eigenstates&EV and uncertainty\\[0.5 ex]
\hline
7& time evolution of wavefunction&-&free particle/math/FT\\[0.5 ex]
\hline
8& EV/&hydrogen atom&energy and momentum/\\[0.5 ex]
& EV and uncertainty&matrix element/eigenstates&math/$\Psi(x,t)$\\[0.5 ex]
\hline
9& FT/Dirac Delta function&math&graphing\\[0.5 ex]
\hline
10& EV/&probability and&free particle/ \\[0.5 ex]
&EV and uncertainty&EV&uncertainty principle\\[0.5 ex]
\hline
11&time dependence of EV&superposition/time dependent&energy and time/ \\[0.5 ex]
& &Schroedinger equation/EV&math/hydrogen atom\\[0.5 ex]
\hline
12& measurement/expansion&collapsed wavefunction/&infinite square well/\\[0.5 ex]
& in eigenfunctions&scalar product/FT&commutation relation\\[0.5 ex]
\hline
13& measurement/&scalar product/&superposition/\\[0.5 ex]
& collapsed wavefunction&eigenvalue&stationary state\\[0.5 ex]
\hline
14& time evolution of wavefunction&hydrogen atom/time dependence&math/time\\[0.5 ex]
\hline
15&-&EV/stationary state/selection&time/time dependent\\[0.5 ex]
& & rules/symmetry (even/odd)&Schroedinger equation\\[0.5 ex]
\hline
16&expansion in eigenfunctions/&stationary state/&infinite square well/\\[0.5 ex]
&time evolution of wavefunction&time dependent function&math\\[0.5 ex]
\hline
17&EV/&symmetry/ &free particle/matrix element/\\[0.5 ex]
& EV and uncertainty&probability and EV&uncertainty relation\\[0.5 ex]
\hline
18&time dependence of EV&EV/superposition&infinite square well\\[0.5 ex]
\hline
19&spin&commutation/uncertainty&math/rotation group\\[0.5 ex]
\hline
20&collapsed wavefunction&hydrogen atom&stationary state/EV\\[0.5 ex]
\hline

\end{tabular}
\vspace{0.1in}
\caption{Examples of categories created for 
each question divided into three groups
with a score of ``5 or 6", ``3 or 4" or ``less than 3". ``EV" is an abbreviation for ``expectation value" and ``FT" is an
abbreviation for ``Fourier Transform".
}
\label{junk2}
\end{table}

\pagebreak

\begin{figure}[h!]
\epsfig{file=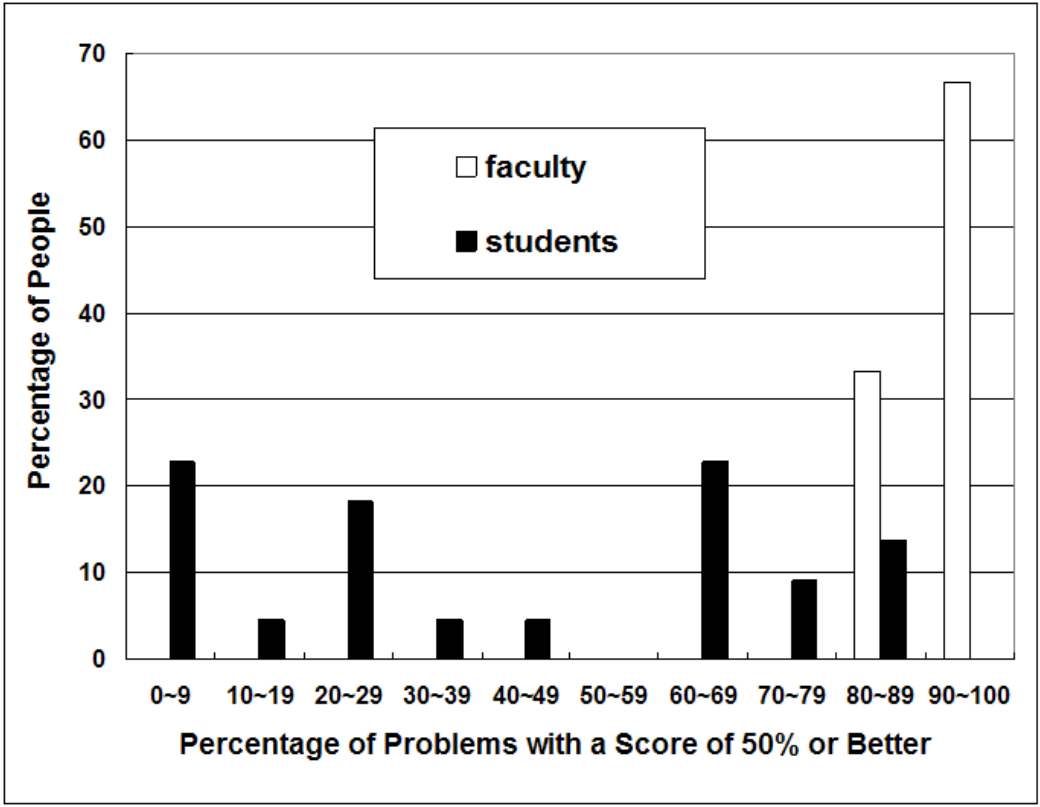,height=4.28in}
\caption{Percentage of people vs. percentage of problems with a score of $50\%$ of better (at least 3 out of 6)}
\end{figure}

\begin{figure}[h!]
\epsfig{file=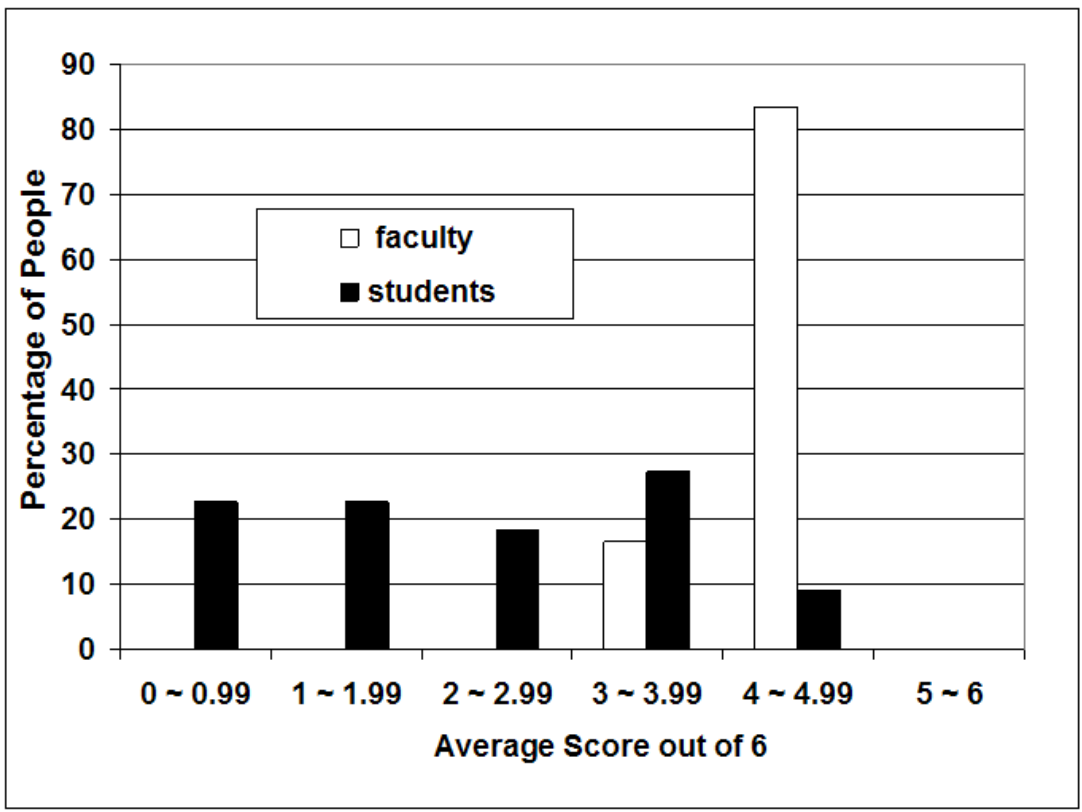,height=4.28in}
\caption{Percentage of people vs. Average score out of 6}
\end{figure}

\end{document}